\documentclass[review,sort&compress]{elsarticle}

\usepackage{amsmath}

\journal{Journal of \LaTeX\ Templates}

\begin{document}

\begin{frontmatter}

\title{A novel method for the line-of-response and time-of-flight reconstruction in TOF-PET detectors 
based on a library of synchronized model signals}

\author[WFAIS]{P.~Moskal}
\author[WFAIS]{N.~Zo\'n}
\author[WFAIS]{T.~Bednarski}
\author[WFAIS]{P.~Bia\l as}
\author[WFAIS]{E.~Czerwi\'nski}
\author[WFAIS]{A.~Gajos}
\author[WFAIS]{D.~Kami{\'n}ska}
\author[WFAIS,PAN]{\L .~Kap\l on}
\author[WCHUJ]{A.~Kochanowski}
\author[WFAIS]{G.~Korcyl} 
\author[WFAIS]{J.~Kowal}
\author[SWIERK]{P.~Kowalski}
\author[WFAIS]{T.~Kozik}
\author[WFAIS]{W.~Krzemie\'n}
\author[WFAIS]{E.~Kubicz}
\author[WFAIS]{Sz.~Nied\'zwiecki}
\author[WFAIS]{M.~Pa\l ka}
\author[SWIERK]{L.~Raczy\'nski}
\author[WFAIS]{Z.~Rudy}
\author[WFAIS]{O.~Rundel}
\author[WFAIS]{P.~Salabura}
\author[WFAIS]{N.G.~Sharma}
\author[WFAIS]{M.~Silarski}
\author[WFAIS]{A.~S\l omski} 
\author[WFAIS]{J.~Smyrski}
\author[WFAIS]{A.~Strzelecki}
\author[WFAIS,PAN]{A.~Wieczorek}
\author[SWIERK]{W.~Wi\'slicki}
\author[WFAIS]{M.~Zieli\'nski}

\address[WFAIS]{Faculty of Physics, Astronomy and Applied Computer Science,
        Jagiellonian University, 30-348 Cracow, Poland}
\address[PAN]{Institute of Metallurgy and Materials Science of Polish Academy of Sciences, \\ 30-059 Cracow, Poland.}
\address[WCHUJ]{Faculty of Chemistry, Jagiellonian University, 30-060 Cracow, Poland}
\address[SWIERK]{\'Swierk Computing Centre, National Centre for Nuclear Research, \\ 05-400 Otwock-\'Swierk, Poland}

\begin{abstract}
A novel method of hit time and hit position reconstruction in scintillator detectors is described. 
The method is based on comparison of detector signals with results stored in a library of synchronized 
model signals registered for a set of well-defined positions of scintillation points. 
The hit position is reconstructed as the one corresponding to the signal from the library 
which is most similar to the measurement signal. 
The time of the interaction is determined as a relative time between 
the measured signal and the most similar one in the library. 
A degree of similarity of measured and model signals is defined 
as the distance between points representing the measurement- and model-signal 
in the multi-dimensional measurement space. 
Novelty of the method lies also in the proposed way of synchronization 
of model signals enabling direct determination of the difference 
between time-of-flights (TOF) of annihilation quanta from the annihilation 
point to the detectors. 
The introduced method was validated using experimental data obtained by means of 
the double strip prototype of the J-PET detector 
and $^{22}$Na sodium isotope as a source of annihilation gamma quanta.
The detector was built   
out from 
plastic scintillator strips
with dimensions of 5~mm x 19~mm x 300~mm,
optically connected at both sides to photomultipliers,
from which signals were sampled 
by means of the Serial Data Analyzer.
Using the introduced method, the spatial and TOF resolution of about 1.3~cm~($\sigma$) 
and 125~ps ($\sigma$) were established, respectively.   
\end{abstract}

\begin{keyword}
\texttt{Scintillator Detectors} \sep \texttt{Positron Emission Tomography} \sep \texttt{Time-of-Flight} \sep \texttt{J-PET}
\end{keyword}

\end{frontmatter}


\section{Introduction}

Positron Emission Tomography (PET) is at present one of the most technologically advanced diagnostics 
methods that allows for non-invasive imaging of physiological processes occurring in a patient's body. 
In the PET tomography the information about the distribution of annihilation points, 
and hence about the density distribution of the administered radiopharmaceuticals inside the patient's body, 
is carried out by pairs of gamma quanta which are registered in detectors surrounding the patient. 
All commercial PET devices use inorganic scintillator materials as radiation detectors 
- usually these are the LBS (BGO) (GE Healthcare), 
LSO (Siemens) or LYSO (Philips) crystals~\cite{Conti2009, Humm2003, Karp2008, Townsend2004}. 
Determination of the interaction point of gamma quanta in PET detectors 
is based on the measurement of charge of signals generated by photomultipliers 
or avalanche photodiodes (APD) connected optically to inorganic crystal blocks 
cut into array of smaller elements. The spatial resolution achievable with this method 
is equal approximately to the dimensions of the small elements of the crystal block. 
Determination of interaction points for both annihilation quanta enables 
reconstruction of the line-of-response (LOR). In turn, the  measurement of the difference between 
the arrival times of gamma quanta to the detectors, 
referred to as time-of-flight (TOF) difference, 
allows to calculate position of the annihilation point along the LOR.  
The TOF resolution of about of 
400 ps achievable with LSO crystals~\cite{Moses1999}, 
allows for a substantial improvement of a signal to noise ratio 
in reconstruction of PET images~\cite{Conti2009, Karp2008, Moses2003}.

Although detectors used in Positron Emission Tomography 
are presently at the highly advanced stage of development 
there is still a large room for improvement, 
and there is ongoing research especially aiming at 
(i) refinement of time resolution by search  
and adaptation of 
new inorganic crystals~\cite{ContiErikson2009, MoszynskiSzczesniak2011, Schaart2010, Schaart2009, Kuhn2006}, 
(ii) reduction of parallax errors due to the unknown depth of interaction (DOI) 
e.g. by application of new geometrical configurations of crystals 
and APD and photomultipliers~\cite{Crosetto2003, Miyaoka1998, Moses1994, Saoudi1999, Vaquero1998}, 
(iii) finding cost-effective solutions which would allow for construction 
of large detectors enabling single-bed whole-body PET imaging 
as e.g. straw tubes drift chambers~\cite{Lacy2001, Shehad2005} 
or large area Resistive Plate Chambers (RPC)~\cite{Belli2006, Blanco2009}, 
and (iv) adaptation of  PET detectors for their simultaneous 
usage together
with MRI and CT modalities~\cite{Crosetto2003, Gilbert2006, Judenhofer2008, Marsden2002, Pichler2008, Quick2011, Townsend2008}.

Recently a new concept of large acceptance Jagiellonian PET (J-PET) system (see Fig.~1.) 
based on strips of polymer scintillators arranged in a large acceptance detectors 
was proposed ~\cite{PCT2014, NovelDetectorSystems, StripPETconcept, TOFPETDetector, JPET-Genewa}.  
The J-PET detector allows to solve the challenges discussed above in an utterly new way. 
It offers improvement of TOF resolution due to the usage of fast plastic scintillators, 
it enables a fusion with MRI and CT modalities in a way allowing 
for simultaneous morphological and functional imaging~\cite{PCT2014-MRI, PCT2014-CT}, 
it permits to determine the depth of interaction~\cite{Smyrski2014}, 
and constitutes a promising solution for single-bed whole-body PET imaging. 
At present it is however in its early stage of development and requires elaborations 
of new hit-position~\cite{NIM_Lech}
and TOF reconstruction methods which would allow to make use of the potential it offers. 
This article is devoted to the presentation of a reconstruction method 
that allows to exploit the advantages of the J-PET detector 
but it may also be applied to other types of scintillator detectors.

In scintillator detectors, amplitude and shape of signals change strongly 
with distance of the hit position to the converter, 
leading to a deterioration of the spatial and time resolution. 
The proposed method of position and time reconstruction turns this disadvantage into an advantage, 
and makes use of the signal shape variation in hit position reconstruction. 
The method is based on determination of the degree of similarity between measured signals 
and standard signals stored in the data base and on a novel concept of signals synchronization.

In the following, for the sake of completeness, the J-PET concept is briefly described. 
Next, in order to facilitate a clear explanation of the reconstruction method we introduce 
a way of representing signals and describe an example of the creation of the library of model signals. 
Further on we describe the invented method of signals' synchronization, which is crucial 
for the reconstruction of LOR and TOF. Finally, the experimental results are presented in the last section of this article.

\section{The J-PET detector system}

\begin{figure}[h!]
\includegraphics[width=0.45\textwidth]{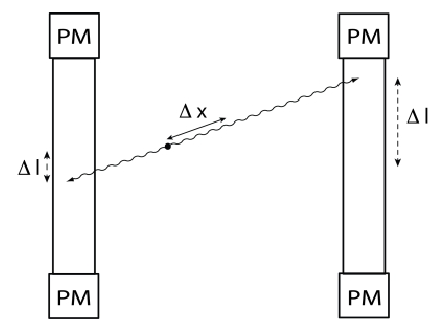} \hspace{0.0cm} \includegraphics[width=0.58\textwidth]{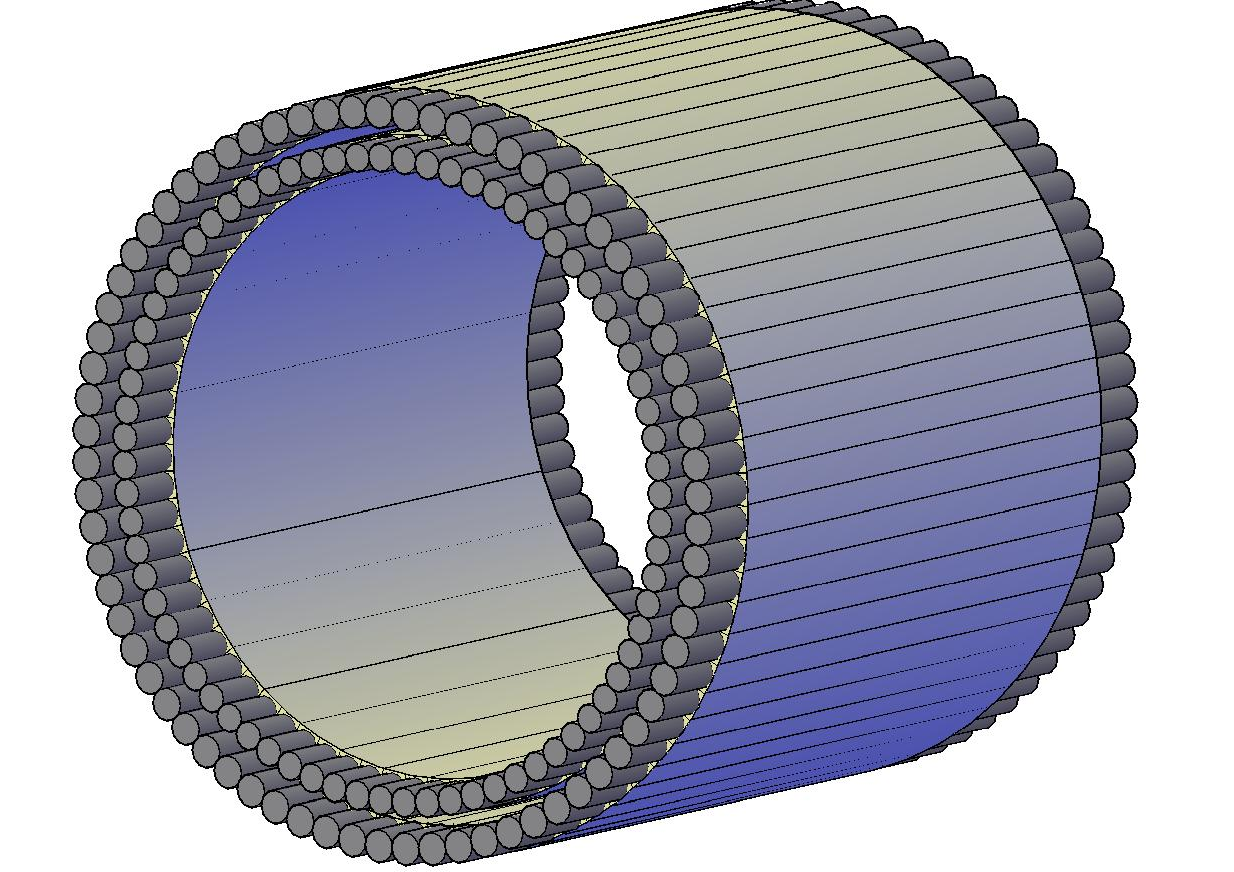}\\
\caption{Left: Schematic view of the two detection modules of the PET detector referred to as J-PET~\cite{NovelDetectorSystems, StripPETconcept}. A single detection module consists of a scintillator strip read out by two photomultipliers labeled with letters PM.  In the first approximation the hit distance from the center of the scintillator ($\Delta l$) is determined based on time difference measured at both ends of the scintillator strip, and the position ($\Delta x$) along the line-of-response is determined from time difference measured between two modules. Right: An example of the two layers version of the J-PET detection chamber.}
\label{uklad}
\end{figure}

The J-PET test chamber is built out of strips of organic scintillator, forming a cylinder.  
One of the possible arrangements of strips is visualized schematically in Fig.~\ref{uklad}.  Light signals from each strip are converted to electrical signals by two photomultipliers placed at opposite ends of the strip. The position and time of reaction of gamma quanta in the detector material can be determined based on the time of arrival of light signals to the ends of the scintillator strips.  In article~\cite{TOFPETDetector} we argued that disadvantages of polymer scintillators due to the low detection efficiency and negligible probability for photoelectric effect can be compensated by a large acceptance, significantly improved time resolution and possibility of usage of several independent detection layers. Especially promising is the possibility of extension of the diagnostic chamber in the J-PET detector which does not entail an increase in the number of photomultipliers and electronics channels when increasing the axial field of view (AFOV). This feature, in contrast to crystal-based PET scanners, allows for building single-bed, whole-body PET scanners without significant increase of costs with respect to scanners with short AFOV. 

The shape (distribution of number of photons as a function of time) and the amplitude of the light signal reaching the photomultiplier changes as a function of the distance between photomultiplier and the place where the light signal was created. Variations of shapes and amplitudes of light signals become stronger with the increasing size of the scintillator, and they constitute a limitation in an achievable time resolution with presently used electronics readout systems utilizing single threshold constant-fraction or constant-level discriminators. In the case of J-PET modality with long polymer scintillator strips, this time resolution determines also the uncertainty of reconstruction of ionization point. Moreover, distribution of amplitude of light signals induced by the gamma quanta is continuous due to the fact that in practice for annihilation quanta only the Compton scattering plays a role in polymer scintillators and the probability for the photoelectric process is negligible. As a consequence the amplitude of signals used for the J-PET image reconstruction varies even if they originate from the same interaction point.  Therefore, a new hit positions reconstruction method is required.

\section{Signal representation}

In the current TOF-PET detectors the reconstruction of line-of-response and of TOF values is based on the charge and time distributions measured for each annihilation event without referring to the external sets of model signals. In this article we present a novel method for the reconstruction of the interaction point in PET detectors and for the reconstruction of  time differences between the arrival of the annihilation quanta to PET detectors. 

\begin{figure}[h!]
\includegraphics[width=1\textwidth]{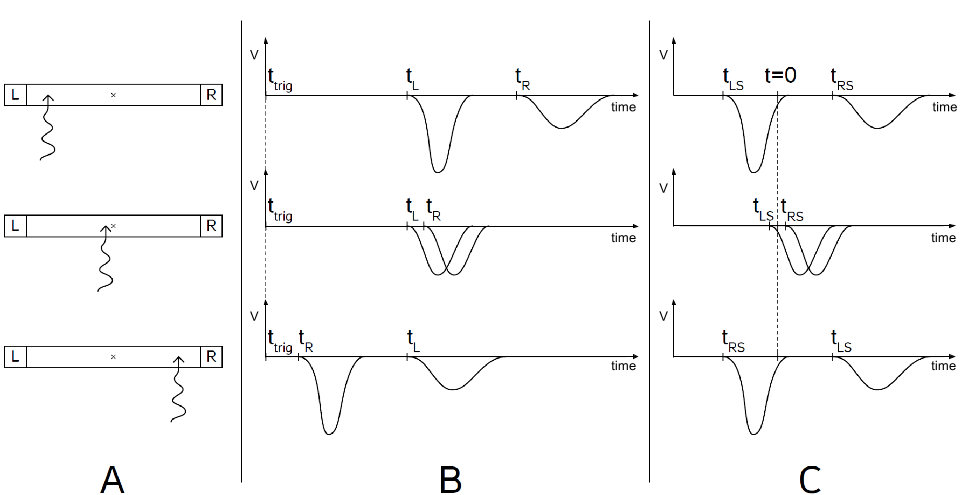}\\
\caption{(A) Pictorial illustration of model signals generation for the J-PET detector. 
(B) Example of generated signals and (C) synchronized signals. Synchronization procedure 
is explained in section 5. Figure (B) shows situation when signal 
from the left photomultiplier was used to define the trigger.  
Letters L and R in figure (A) denote left and right photomultiplier, respectively. 
The time of the trigger is denoted by $t_{trig}$, and beginnings of signals 
for left and right side are indicated by $t_L$ and $t_R$, respectively. 
In figure (C) index “s” is added to indicate times after synchronization. More details are given in the text.}
\label{fig2}
\end{figure}

The description of the proposed method is based on the example of the J-PET detector (Fig.~\ref{uklad}), 
consisting of scintillator strips connected optically at two ends to photomultipliers. 
Fig.~\ref{fig2}B shows schematically signals (voltage as a function of time) from a single detection 
module for a few irradiation positions indicated in Fig.~\ref{fig2}A. Figs.~\ref{fig2}A and~\ref{fig2}B 
illustrate qualitative changes of the signal shape and amplitude as a function of the hit position 
of the gamma quantum along the scintillator strip. 

Signals from photomultipliers are processed by the read-out electronics enabling determination of their charges and times at which they pass through given reference voltages~\cite{Palka2014}.

The method described below may in general be used in PET modalities in which signals are sampled in the voltage domain by means of multi-threshold constant-level discriminators, or in the domain of the fractions of amplitude by means of constant-fraction discriminators. Preferably both kinds of discriminators should be applied for sampling since they deliver complementary information. Constant level discriminators are used to determine a moment of time in which the detector signal crosses a defined reference voltage, whereas the constant fraction discriminators allow to determine the time when signal crosses voltage level equal to a certain fraction of the signal’s amplitude. 

The shape and amplitude of signals corresponding to the registration of gamma quantum in the scintillator detector changes from event to event and depends on many factors such as e.g. statistical character of light emission, energy absorbed in the scintillator, the location of the interaction point in the scintillator strip and many others. Therefore, in order to reach high precision of the time and position measurement, signals in the J-PET detector are split and read by means of the multi-threshold constant level and multi-threshold constant fraction discriminators.

The registration of each gamma quantum may be represented as a point in a measurement space $\Omega_m$ with the number of dimensions equal to the number of measured parameters such as times or charges of signals induced in a single detection module. Each time or charge measurement increases the $\Omega_m$ space by one dimension. Further on we denote number of measurements done on signals induced by a single gamma quantum by 
$N_m = 2 (N_f+N_{cr}+N_{cf}+1)$, where $N_f$ denotes the number of thresholds at the constant fraction discriminators, $N_{cr}$ is the number of time measurements with a constant-level discriminator at the leading edge of the signal, $N_{cf}$ stands for the number of measurements with a constant level discriminator at the trailing edge, and $(+1)$ corresponds to the charge measurement. Factor of „2” before the parenthesis reflects the fact that each scintillator is read out by two photomultipliers. Hence, a result of the registration of a single gamma quantum corresponds to a point $P$ in an $N_m$ dimensional measurement space $\Omega _m$. First $N_m/2$ coordinates of the point correspond to the measurement at one side of the strip and the next $N_m/2$ coordinates to the measurements at the other side. 

Various coordinates of the point and their mutual relations are sensitive in a different way to the changes of amplitude, time or the shape of the signal. Therefore, based on the measured signals, it is possible to disentangle information of the time, position and energy deposited in the scintillator.  For example, values of $P(i)$, for $i=1,... ,N_f$, corresponding to the measurement of the time by the constant-fraction discriminators, are sensitive to variation of signal’s shape, but are not sensitive to the changes of the signals amplitude provided that the shape of the signal and its time of origin are not changing. On the other hand values of $P(i)$ corresponding to the measurement of time with the constant-level discriminators depend on the signal amplitude even if the shape and time of the origin of the signal are not changing. Moreover, time differences measured for a given reference voltage at different sides of the scintillator strip strongly depend on the place of the gamma quantum interaction.  In general, coordinate $P(i)$ representing the time measurement may be expressed as:
$P(i) = t_{measurement}(i) + t_{delay}(i) + t_{trig}$, 
where $t_{trig}$ acquires the same value for all coordinates $i$, $t_{measurement}(i)$ denotes the time at which signal crosses a reference voltage at discriminator corresponding to the ith dimension in the space $\Omega_m$, and $t_{delay}(i)$ stands for the constant which is subject to calibration, and which denotes the time elapsed to the moment of the measurement from the moment at which electronic signal would be created if it was created at the edge of the scintillator without delays due to the photomultiplier, cables and read-out electronics. The time offsets $t_{delay}(i)$ may be determined for each detector module with respect to the reference detector utilizing “beta plus” radioactive isotopes rotating inside a scanner  (see e.g. Fig.~\ref{fig3}) or by other methods~\cite{Silarski2014}. Therefore, we assume that the $t_{delay}(i)$ constants are known for each detection module and for simplicity, and without loss of generality, we will skip them in the further considerations.

\section{Generation of the library of model signals}

\begin{figure}[h!]
\includegraphics[width=1\textwidth]{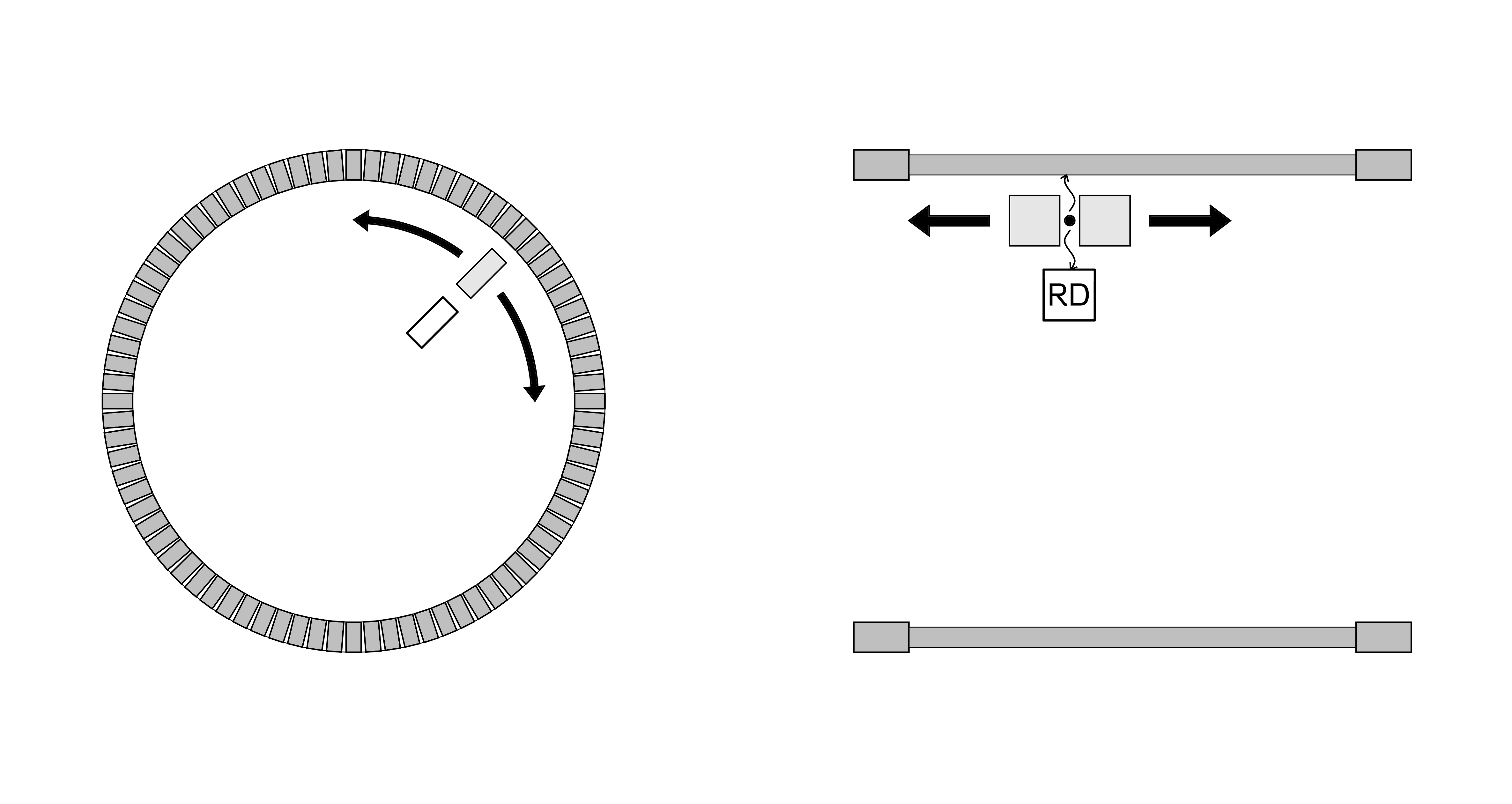}\\
\caption{Schematic cross sections of the J-PET detector with a collimated beam of annihilation quanta rotating around the axis and moving along the detector, as indicated by the arrows. RD denotes a reference detector permitting to select signals corresponding to the annihilation quanta and improving the collimation.}
\label{fig3}
\end{figure}

The reconstruction method described in this article requires generation of a data base of synchronized model signals for various interaction points. The library of model signals is generated by scanning the scintillator strip with a collimated beam of annihilation quanta with profile smaller than the spatial resolution required for the hit position reconstruction. For example a beam with the profile width of FWHM equal to 1~mm can be used. Scanning may be performed using a source of annihilation gamma quanta placed inside a collimator which may rotate around the axis of the detector and which can be moved simultaneously along the strips, giving possibility to irradiate each place of the detector's inner surface as it is depicted in Fig.~\ref{fig3}. Movements of the collimator must be synchronized with the data acquisition system in order to assign a place of irradiation to each measured signal. The information about the position of irradiation is added to each signal in the library.  For each position a high statistics of signals is collected.

As a next step, signals in the generated data base are synchronized in such a way that for all of them a time corresponding to the reaction of the gamma quantum amounts to zero, as it is demonstrated in Fig.~\ref{fig2}C. 
The synchronized library of model signals constitutes a set of baseline points in the space $\Omega_m$, and it is used in the reconstruction procedure for determining sets of LOR and TOF values from the PET measurements. The synchronization procedure is described in the next section. 

\section{Method of the model signals synchronization}

A raw data base determined as described in the previous section constitutes a set of points $P$ in a space $\Omega_m$. 
Coordinates of points $P$ correspond to moments of time in which signals pass the discriminator 
thresholds with respect to the time of the triggering signal. 
Few exemplary signals from the raw data base are shown in Fig.~\ref{fig2}B.  
The signals may be synchronized with respect to the signal of the reference detector, 
which allows for the precise determination of the calibration offsets ($t_{delay}$), 
since in this case an average value from the large statistics sample of signals can be used. However, the reference detector introduces smearing of time when the single signals from the library are used for the reconstruction of time of the interaction. Therefore, we introduce a synchronization technique independent of the performance of the reference detector.
The main idea of synchronization of signals in the data base lies in shifting them in time such that the moment when gamma quantum hits the detector would be the same for all events in the library. The absolute value of this time is not relevant, therefore for simplicity the signals in the library will be synchronized to the hit-time equal to zero. 
With the appropriate choice of the calibration constants, 
the time at which gamma quantum undergoes scattering 
in the detector with respect to the time of the trigger signal is determined  by the mean value: $t_{hit} = (t_L + t_R)/2$, where $t_L$, and $t_R$ denote the beginning of the signal measured at the left and right side of the strip, respectively. The beginning of the signal, represented by a given point P may be determined based on the coordinates of this point by fitting to them a function which describes the shape of the signal at its leading edge. A function describing the shape of signals for a given hit-position may be determined experimentally e.g. by averaging signals in a high statistical sample collected at this position. 
Synchronization is realized by transforming point $P$ into point $P_s$  using the following prescription: $(P_s(i) = P(i) + t_{synch}, i=1,... ,N_m)$, where a value of $t_{synch}$, is chosen such that after the transformation $t_{Ls} + t_{Rs} = 0$.  Thus, $t_{synch} = - (t_L + t_R)/2$, and it needs to be determined separately for each point from the data base. An example of synchronized signals is shown in Fig.~\ref{fig2}C. Note that synchronization of signals in the library to the value of $t_{hit} = 0$ implies that always one of the times from the pair $(t_{Ls}, t_{Rs})$ is negative and the other positive. The above described synchronization procedure enables to determine not only LOR but also TOF for each registered event.

\section{The reconstruction method of LOR and TOF}

The reconstruction of the time and position of the interaction of gamma quanta in the detector is based 
on comparison of measurement signals for a given event with synchronized model signals stored in the library. 
The hit position is reconstructed as this which corresponds to the signal from the library which is most similar 
to the measurement signal, and hit time of the interaction is reconstructed as a relative time between the measured 
signal and the most similar one in the library. A degree of similarity is defined as a distance between points 
representing the measurement- and model-signal in the multi-dimensional measurement space $\Omega_m$. 
The distance is determined taking into account measurement uncertainties of charge and times at various reference 
voltages and correlations between these measurements.

\begin{figure}[h!]
\includegraphics[width=1\textwidth]{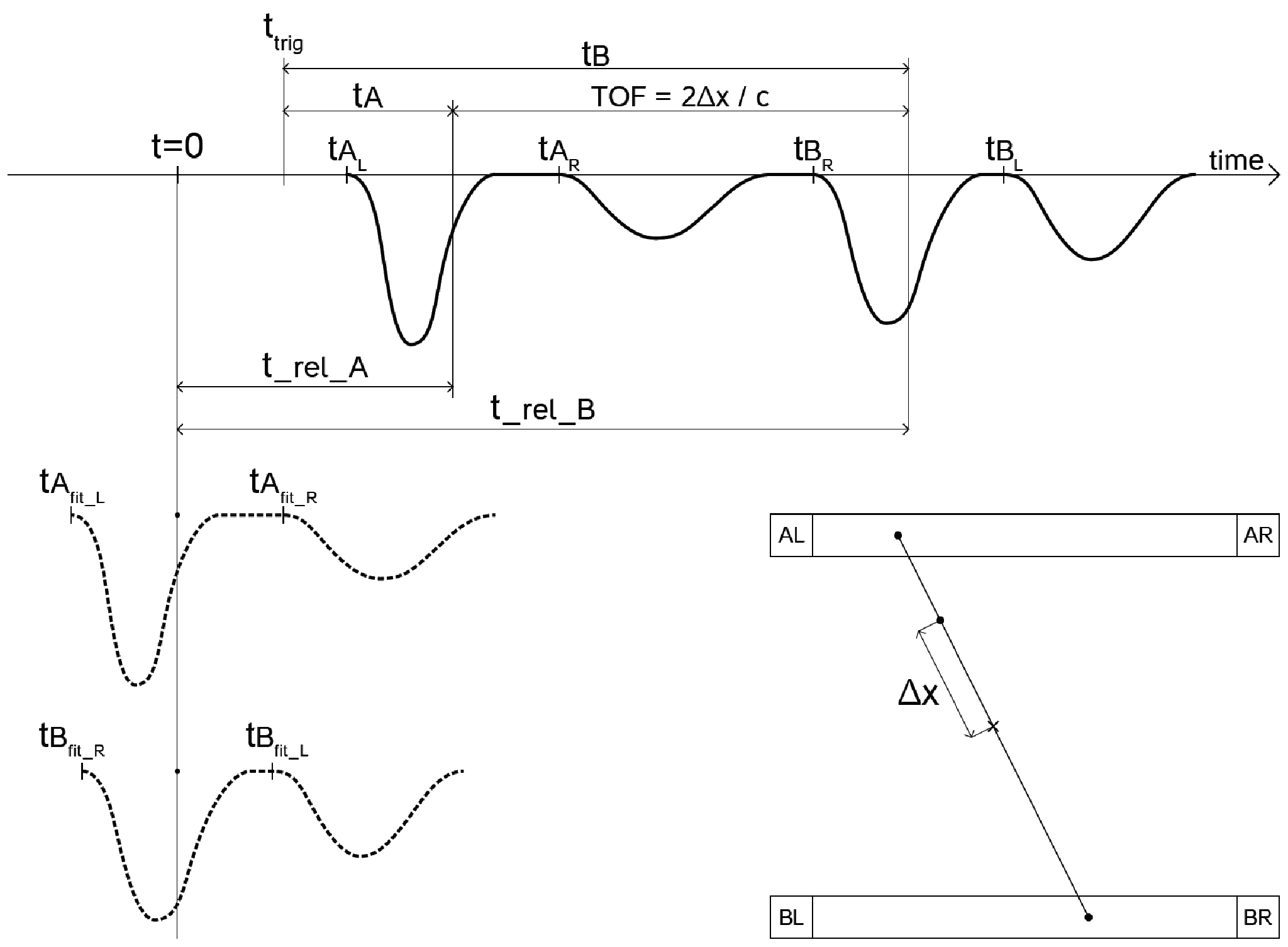}\\
\caption{Pictorial illustration of the method for TOF reconstruction based on the comparison of the measured signal (solid line) with the synchronized model signals (dotted lines). For the detailed description see text. $tA_L$, $tA_R$ denote the beginning of signals measured at the left and right side of the detector A, and $tA$ stands for the hit-time calculated as a mean value of times $tA_L$ and $tA_R$. Analogously, beginnings of signals from detector B and signals from the library are shown.  In the right-lower corner of the figure, an event corresponding to the measurement signal is illustrated. AL, AR, BL, BR denote photomultipliers on the left and right sides of the A and B detectors, respectively.  A cross indicates a center of the line-of-response, and the dot on the line denotes the point of annihilation which is by $\Delta x$ away from the center of the LOR.}
\label{fig4}
\end{figure}

In order to determine the time and position of the gamma quantum interaction in a given scintillator strip, 
the algorithm searches through the set of points in the library of synchronized model signals to find a point $P_{s0}$ which is closest to the point $P$ representing the measurement signal. 
Measurements, and as a consequence coordinates of points in the measurement space $\Omega_m$, 
are burdened with uncertainties which may be correlated with each other. 
These uncertainties are described by the covariance matrix which should be determined 
for each detection module separately. An inverse covariance matrix constitutes a metric determining distance 
in the measurement space $\Omega_m$. Such defined distance, which takes into account measurement uncertainties 
and their correlations is in the literature referred to as Mahalanobis distance~\cite{Mahalanobis1936}. 
In general a measure of the distance between points, and thus the measure of the degree of similarity between 
signals represented by these points, may be defined in many manners, as for example: (i) probability that two compared 
signals are the same (applicable in the case of the maximum likelihood method), 
(ii) Chi-square ($\chi^2$) value used in the case of the minimum square method, or 
(iii) Hausdorff distance used as a degree of resemblance between two signals~\cite{Huttenlocher1993}. 
In order to compare a measurement signal (from the diagnosis of the patient) represented by point $P$ 
with the model signal from the synchronized library represented by point $P_s$ one has to perform minimization 
of the distance between points $P$ and $P_s$ varying the relative time ($t_{rel}$) between the synchronized 
basis and the signal $P$ from the diagnosis of the patient. This first step may be understood as superimposing 
of signals $P$ and $P_s$ on each other. Thus, the degree of similarity, e.g. a Mahalanobis distance between 
points $P$ and $P_s$ is expressed as a function of $ t_{rel} $: $Mahanalobis(P+T_{rel}, P_s)$, 
with $T_{rel} = (t_{rel},... ,t_{rel})$, where $t_{rel}$ is a fit parameter. 
For each point $P_s$ from the synchronized library a minimum value of $min[Mahalanobis(P+T_{rel}, P_s)]$ 
is determined with respect to $t_{rel}$, and next as a point $P_{sfit}$, 
being closest to the point $P$, such point $P_s$ is chosen, 
for which a value of $min[Mahanalobis(P+T_{rel}, P_s)]$ is the smallest.  
Finally, a point of interaction of the gamma quantum is determined as a place at which 
a beam of annihilation quanta was directed at the moment when a signal represented 
in the library by $P_{sfit}$ was registered, and as the time of the interaction the value 
of $ t_{rel} $ is chosen for which $Mahanalobis(P+T_{rel}, P_{sfit})$ is minimal. 
Such choice of the value of time of interaction constitutes one of the crucial 
ideas of the described reconstruction method. It ensures that the difference between 
times of interactions reconstructed in different detectors for the quanta 
from the same annihilation process, correspond to the true difference (TOF) between 
times of arrival of these quanta to the detectors. 
This feature is proven below by the reasoning illustrated in Fig.~\ref{fig4}. 

In order to focus the attention of the reader, without loss of generality, we assume that the gamma quanta were registered in detectors A and B (see right-lower corner of Fig.~\ref{fig4}). Then, $tA$, a time of the reaction of gamma quantum in the detector A with respect to the time of the trigger, may be determined as: 

$tA = (tA_L + tA_R ) / 2$,  

and analogously:

$tB = (tB_L + tB_R) / 2$,

where $tA_L$ denotes the time of the beginning of the signal generated in the left photomultiplier of detector A measured with respect to the time of the trigger, and $tA_R$, $tB_L$ and $tB_R$  denote correspondingly  beginning of signals in the right side of detector A and beginning of signals in left and right sides of detector B.
Solid line in Fig.~\ref{fig4} represents pulses registered in left and right side of detectors A and B for an exemplary event where annihilation process occurred by $\Delta x$ away from the center of LOR. The dotted lines indicate signals from the synchronized library of model signals which were determined by means of reconstruction procedure as most similar to the measurement signal indicated by solid line.  The reconstruction procedure described above returns $t_{rel}A$ and $t_{rel}B$ as times at which gamma quanta hit detector A and B, respectively. Thus, Fig.~\ref{fig4} clearly shows that: 
 
$t_{rel}A = tA + t_{trig}$ and $t_{rel}B = tB + t_{trig}$, 

and hence

$t_{rel}B - t_{rel}A = tB - tA = TOF$. 

It is important to note that the above result is independent of the time of the trigger. The result of the above reasoning proves that the synchronization and reconstruction methods presented in this article allows for the direct determination of LOR and TOF once the most similar signal to the measurement signal was found in the library of synchronized model signals.

\section{Double-strip J-PET prototype}
\begin{figure}[h!]
\includegraphics[width=1\textwidth]{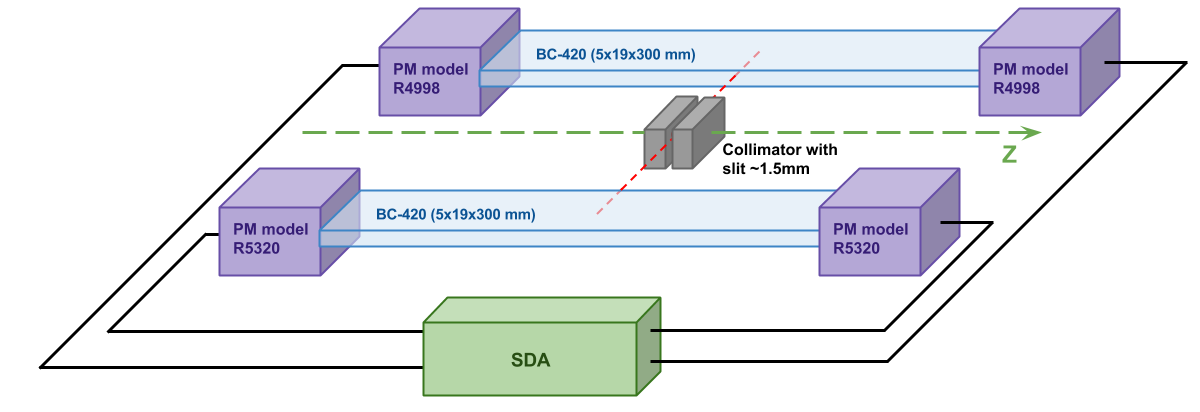}\\
\caption{A schematic view of the double-strip J-PET prototype. Detailed description is given in the text.}
\label{figSetup}
\end{figure}

\begin{figure}[h!]
\includegraphics[width=1\textwidth]{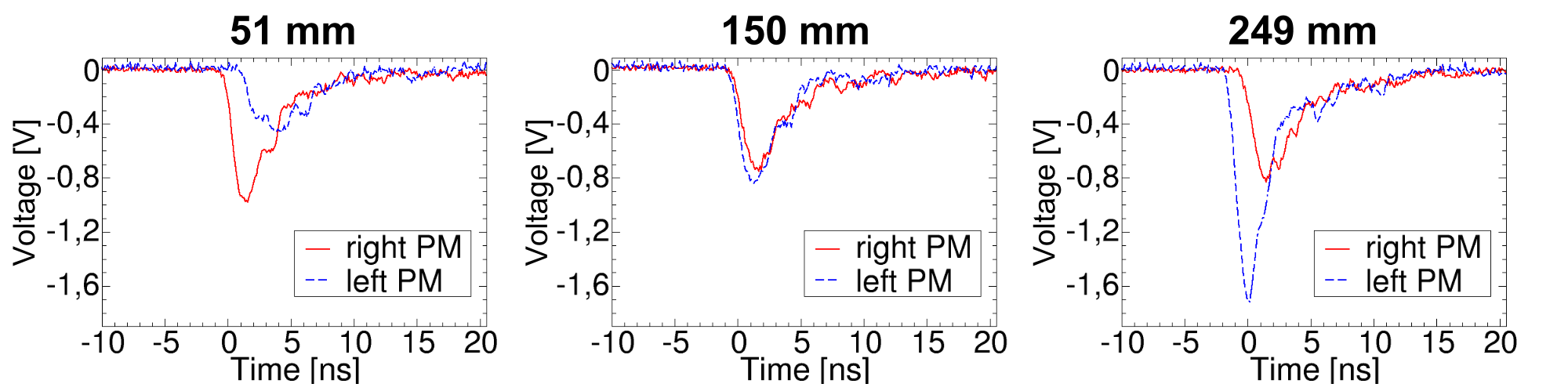}\\
\caption{Exemplary signals measured closer to the left photomultiplier at positions z~=~51~mm (left), 150~mm (center) 
         and closer to the right photomultiplier at position z~=~249~mm (right).
\label{fig3Signals}
        }
\end{figure}

\begin{figure}[h!]
\includegraphics[width=0.5\textwidth]{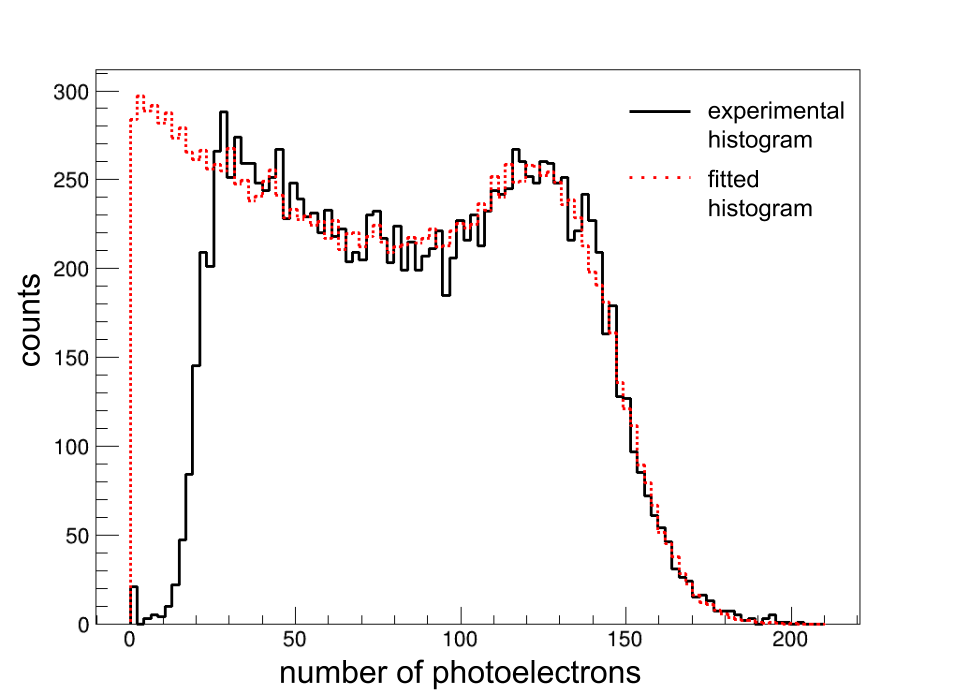}
\hspace{0.0cm} 
\includegraphics[width=0.5\textwidth]{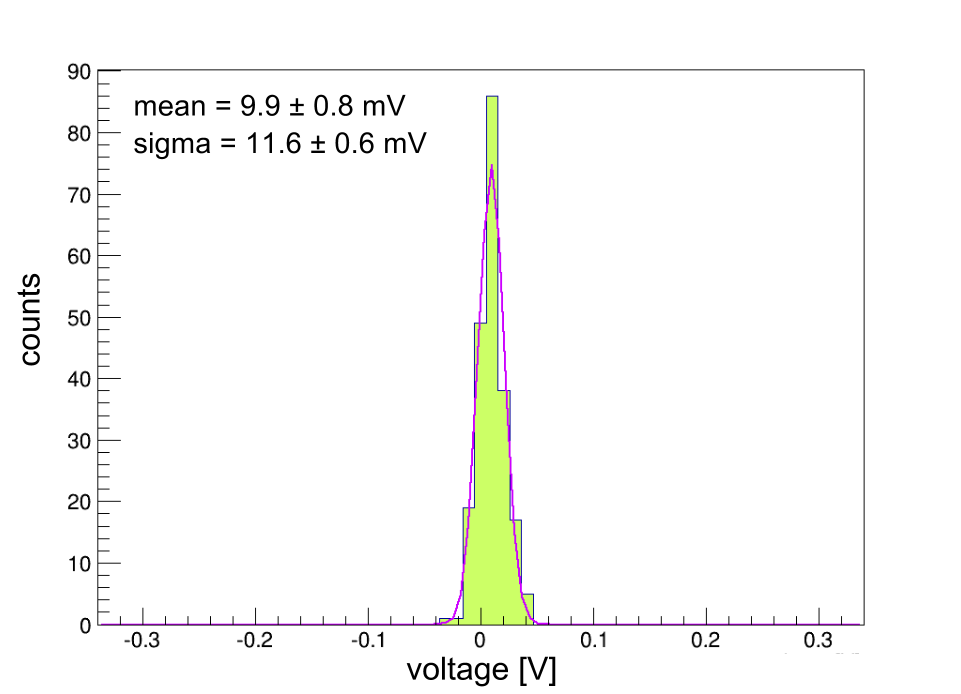}\\
\caption{Left panel: (Solid black histogram) Charge distribution of signals from one of the photomultipliers
determined by irradiating the center of the scintillator 
strip with collimated beam of annihilation gamma quanta. 
(Dashed red histogram) Theoretical distribution of energy of scattered electrons~\cite{Klein} 
convoluted with the resolution of the detector~\cite{NIM-1strip} and fitted to the experimental data with normalization and energy calibration constants 
as the free parameters. Details of the fitting procedure are described in reference~\cite{NIM-1strip}.
The charge of the signal is expressed in the number of photoelectrons estimated
using a method described in reference~\cite{BednarskiBAMS}.
The lower range of the experimental spectrum is cut by the threshold set at SDA.
Right panel: An example of the distribution of the noise of the measured signals.
The figure shows distribution of voltage of a single signal
for times lower than -2.5 ns, i.e. before the onset of real pulse (examine Fig.~\ref{fig3Signals}).
A superimposed line indicates a result of the fit of the Gaussian distribution. 
\label{figHistogramFit}
}
\end{figure}

The J-PET detector system shown in Fig.~\ref{uklad} is axially symmetric and its performance may be 
tested using a double strip prototype which allows 
for simultaneous registration of two annihilation quanta and reconstruction 
of both LOR and TOF. 
Therefore, the functioning of the J-PET detector 
and validation of the reconstruction method proposed in this article was verified
using the double strip prototype outlined in Fig.~\ref{figSetup}.
The prototype is built out from BC-420 scintillator strips~\cite{SaintGobain}
with dimensions of 5~mm x 19~mm x 300~mm wrapped with the 3M Vikuiti specular reflector foil~\cite{3M}. 
The strips are read out at both sides by Hamamatsu R4998 and R5320 photomultipliers~\cite{Hamamatsu}. 
Two different kinds of available photomulitpliers R4998 and R5320 were used. However, they differ only in quantum efficiency
for the registration of photons in the ultra-violet region not relevant for the emission spectra of BC-420 scintillator.
The source of $^{22}Na$ with its active part in the form of cylinder
with diameter of 3~mm and thickness of 1~mm was 
located within a lead collimator with a 1.5~mm wide and 20~cm 
long slit providing a well collimated beam of annihilation quanta with the spatial profile of about 1.5~mm~(FWHM).
A dedicated mechanical system and a step-motor allowed to 
move the collimator along "z" axis with the precision of a fraction of millimeter
so as to permit irradiation of chosen point within the detector. 
Signals from photomultipliers were sampled with 100 ps intervals by means of the Serial Data Analyzer (Lecroy SDA6000A).
A library of model signals was created by moving the collimator in steps of 3~mm and 
collecting 5000 events for each irradiated position.
The information about the position of irradiation is added to each signal in the library.  
Exemplary signals measured for three different positions are shown in Fig.~\ref{fig3Signals}.
Left and right panels of Fig.~\ref{fig3Signals}
present signals measured closer to the left and right photomultiplier, respectively. 
As expected, signals measured by the photomultiplier nearer to the interaction site
are larger and arrive earlier than signals from the other photumultiplier, whereas signals collected in the center of the strip
are characterized by the same shape and the same onset time (the signals in the center may not ideally overlap
due to the possible differences in the photomultiplier transit times and different delays of the SDA channels). 
Plastic scintillators such as BC-420 consist of carbon and hydrogen and due to the low atomic number of these elements
the probability for the photoelectric effect for the 511 keV annihilation quanta is negligible.
In practice interactions of annihilation quanta in plastic scintillators
occur only via Compton scattering~\cite{BAMS-Szymanski,NIM-1strip}, 
and the spectrum of energy deposition and hence distribution of charge 
of registered signals is continuous and ranges from zero to 0.341~MeV (2/3 of electron mass).  
The example charge spectrum of signals registered by irradiating the middle of the scintillator
is shown as a black solid line in the left panel of Fig.~\ref{figHistogramFit}. 
In order to avoid large fluctuations in shape of signals consisting of small numbers
of photoelectrons for the further analysis we have selected only these events for which  
energy depositions were larger than 0.2~MeV in both scintillator strips.  
In order to find relation between the measured charge
and deposited energy
the Klein-Nishina formula~\cite{Klein} convoluted with 
the detector resolution was fitted to the experimental data
with energy calibration constant 
and normalisation as free parameters~\cite{NIM-1strip}. 
An example of result of such fit is shown as dashed red histogram in Fig.~\ref{figHistogramFit}.
It is worth to stress that in a reconstruction of the tomographic image such filtering of
signals will be performed and only 
signals with energy deposited larger than  0.2~MeV
will be considered in order to suppress events originating from scattering
of the annihilation quanta in the patient's body~\cite{TOFPETDetector}.

\subsection{LOR and TOF reconstruction}
According to the description included in section 5 we have synchronized model signals in the library
such that the time of the interaction of gamma quantum corresponding to each signal is equal to zero. 
To this end each model signal was shifted in time by the value of 
$t_{synch} = - (t_L + t_R)/2$. For the sake of simplicity, 
we have determined $t_L$ and $t_R$ as times
at which the signal cross the threshold voltage of 80~mV. 
A value of 80~mV was chosen to optimise smearing of time due to the noise and due to the time walk effect.
In order to decrease the influence of the time walk effect the threshold should be as low as possible but
on the other hand it should be sufficiently high 
to decrease the influence of the electronic noise which typically amounts to about 10~mV to 20~mV(sigma) depending on the applied voltage.
An example of the electronic noise spectrum is presented in Fig.~\ref{figHistogramFit}. 
It is also important to stress that before the determination of $t_L$ and $t_R$ each signal was corrected for the 
pedestal which in the example shown in Fig.~\ref{figHistogramFit} amounts to 9.9 mV.

For the test of the reconstruction method  introduced in section 6 
we have chosen events measured when the collimated beam was irradiating strips at the following positions: 
z~=~51~mm, z~=~99~mm, z~=~150~mm, z~=~201~mm and z~=~249~cm.
The position and TOF resultions of the J-PET prototype are determined 
from distributions of the differences between true 
and reconstructed values of position and TOF, respectively. As  $true$ positions we denote real positions
of irradiation. 

For the purpose of this demonstrative analysis each signal is represented as an array 
with coordinates P(i), where i~=~1,...,42. According to the description from section~3 
first 21 coordinates describe results of measurement with left signal and next 21  corresponds to the right signal. In particular:

P(1),...,P(10) corresponds to times at which a left signal is crossing ith threshold voltage $V_i$ = 60 mV + (i-1) $\times$ 50 mV, 

P(11),...,P(20) corresponds to times at which a left signal is crossing ith fraction of its amplitue $f_i$ = 0.1 + (i-11) $\times$ 0.05,

P(21) corresponds to the signal charge. And analogously P(22),...,P42 are defined for the right signal. 

\begin{figure}[h!]
\centering
\includegraphics[width=0.6\textwidth]{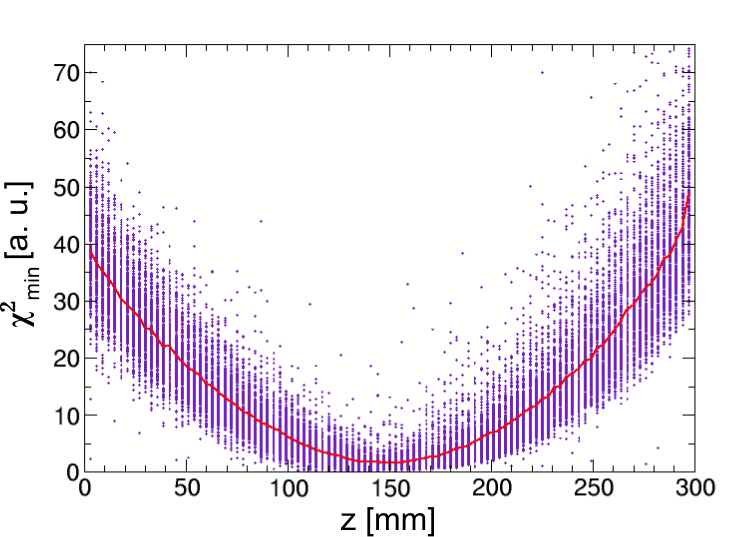}\\
\caption{
$\chi^2_{min}$ as a function of position assigned to the model signals. 
Superimpsed solid line indicates average value of $\chi^2_{min}$ determined for each position separately. 
\label{Chi2}
}
\end{figure}

As introduced in section 6, $P$ and $P_{s}$ denote signals from the tested subset (P) and from the library 
of synchronized model signals ($P_{s}$), respectively.  
In order to reconstruct a place of gamma interaction corresponding to a given signal P,
this signal is compared with all signals in the library.  Next, position 
assigned to the most similar model signal 
is taken as the reconstructed position. 
As a measure of similarity a $\chi^2$ like variable is used
which is defined as follows:
\begin{equation}
\begin{array}{c}
 \chi^2(P,P_s,t_{rel}) = \\
 \sum_{i=1}^{20}(P(i) - P_s(i) - t_{rel})^2/\sigma^2(t) \\ 
 +~\sum_{i=22}^{41}(P(i) - P_s(i) - t_{rel})^2/\sigma^2(t) \\
 +~(P(21) - P_s(21))^2/\sigma^2(Q)\\  
 +~(P(42) - P_s(42))^2/\sigma^2(Q), 
 \end{array}
\end{equation}
where $\sigma(t)$~varies between $\sim 13$~ps and $\sim 40$~ps depending on the thereshold,
and it was determined by the measurement of distributions of time differences 
of the same signal split into two  
different SDA channels.
As regards the charge, 
the studies described in reference~\cite{NIM-1strip} revealed that the uncertainty of the measurement of a signal's charge is dominated by the statistical
fluctuation of the number of photoelectrons. Therefore, we express charge in units of photolectrons $N_{phe}$ and estimate its uncertainty as $\sqrt{N_{phe}}$.
Thus, the $\sigma^2(Q)$ denoting the variance of the difference of the measured charges is equal to the sum of the number of photoelectrons from the compared signals. 
Fig.~\ref{Chi2} presents an example plot 
of minimum values of $\chi^2$ determined during the reconstruction process for one of the P signals measured at position z~=~150~mm.
Each point at this plot corresponds to a minimum value of $\chi^2$ 
resulting from the comparison 
of signal $P$ with a model signal $P_{s}$.
A minimum value of $\chi^2$ is found with respect to $t_{rel}$.
\begin{figure}[h!]
\includegraphics[width=0.45\textwidth]{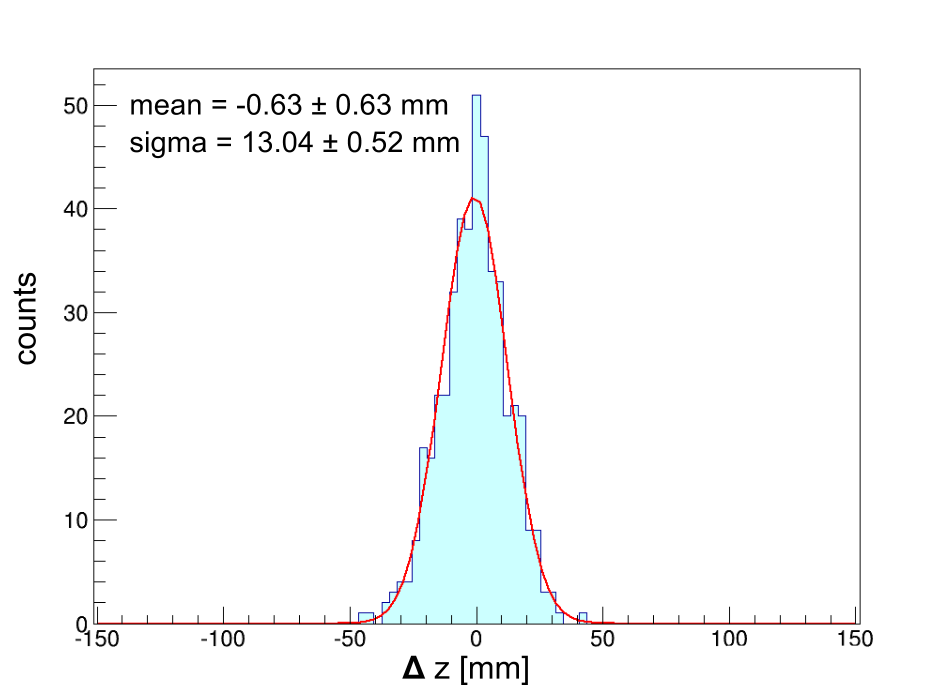}
\hspace{-0.0cm} 
\includegraphics[width=0.45\textwidth]{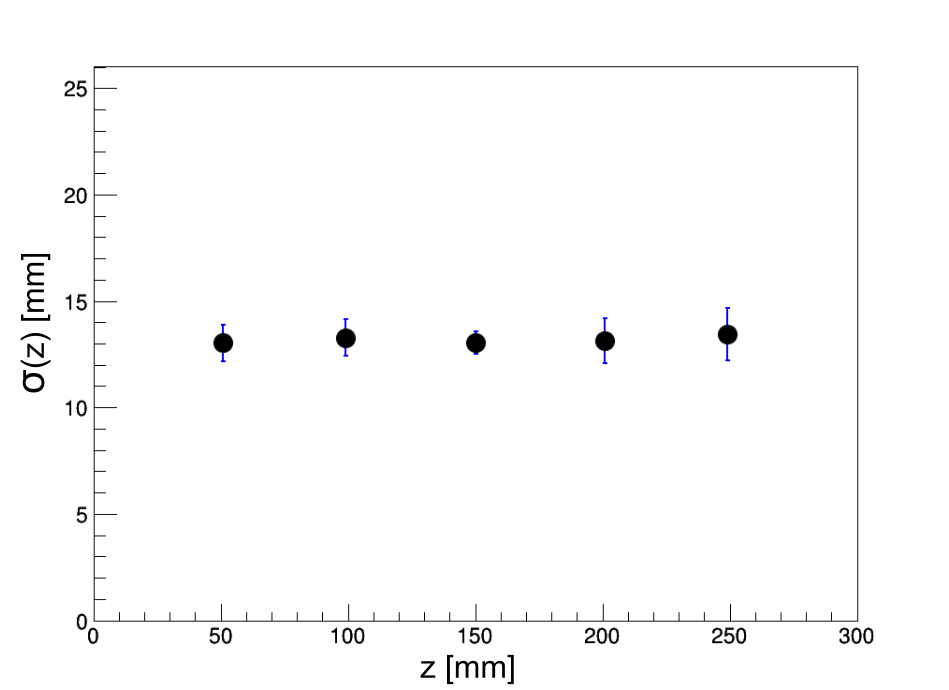}\\
\caption{
(Left) Distribution of differences between the true and reconstructed position for signals measured at z~=~150~mm.
(Right) Position resolution as a function of the place of gamma quantum interaction.
\label{figPositionResolution}
}
\end{figure}
It should be noted that for each position  many signals in the library are stored, and hence many points 
are visible in Fig.~\ref{Chi2} at each position. 
Finally, as a most similar signal to the processed $P$ signal, 
such model signal $P_{sfit}$ is chosen for which $\chi^2_{min}$ is the smallest.
Position assigned to $P_{sfit}$ 
is taken as reconstructed position corresponding to signal $P$. 
Left panel of Fig.~\ref{figPositionResolution} 
shows distribution of differences between the true and reconstructed position for 
signals measured 
at z~=~150~mm, where by fitting a Gaussian function the resolution of $\sigma$~$\approx$~13~mm was established.
Right panel indicates that this resolution does not change with the position. 
\begin{figure}[h!]
\includegraphics[width=0.45\textwidth]{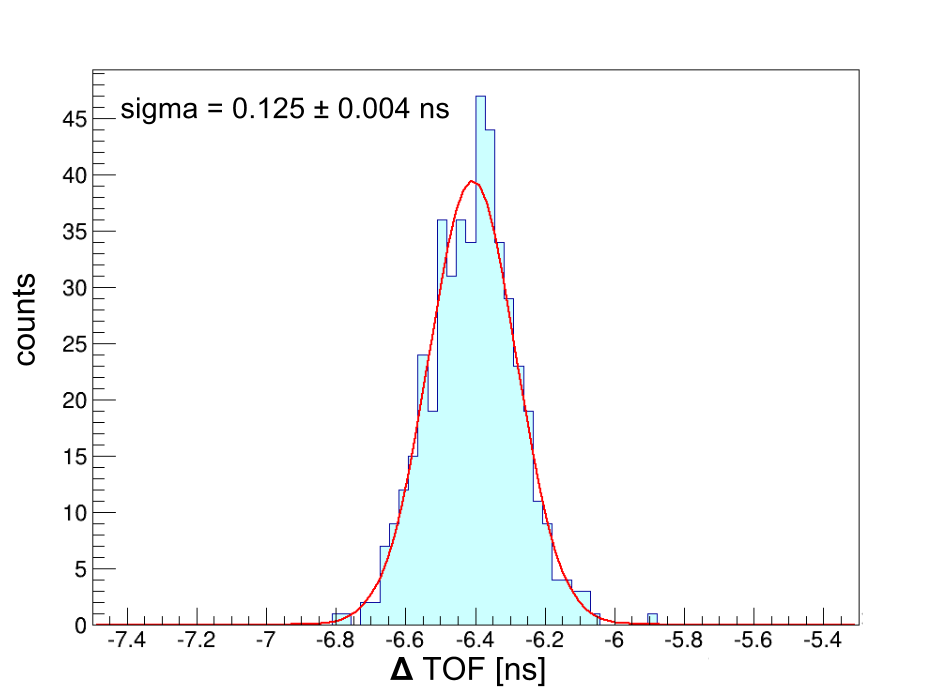}
\hspace{-0.0cm} 
\includegraphics[width=0.45\textwidth]{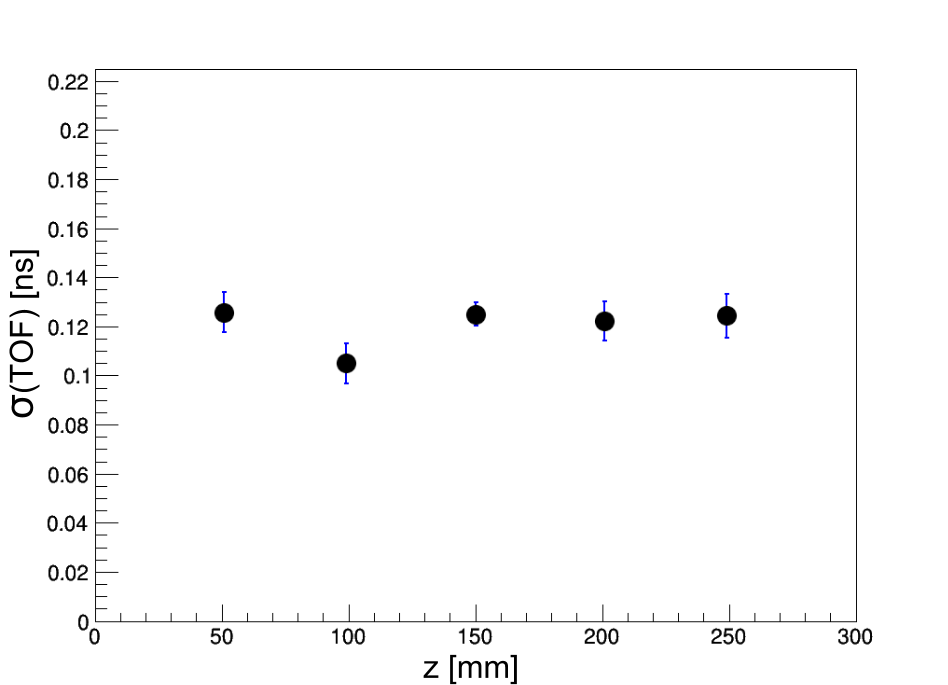}
\caption{
(Left) Distribution of differences between the true and reconstructed TOF for signals measured at z~=~150~mm.
The non-zero value of the mean is mainly due to the difference in the cable lengths and it is matter of calibration. 
(Right) TOF resolution as a function of the place of gamma quantum interaction.
}
\label{figTOFResolution}
\end{figure}
As the last step of the analysis a
TOF value is reconstructed as $t_{rel}A$ - $t_{rel}B$
according to the procedure described in section~6 and illustrated in Fig.~\ref{fig4},
where $t_{rel}A$ and $t_{rel}B$ stand for the $t_{rel}$ values for which the 
global minima of $\chi^2$ were found for first and second detector strip, respectively. 
As regards the true TOF value, it is equal to zero since the source was positioned
in the middle between the detection modules. However, due to the differences in delays caused 
by different electronics channels and cables, the reconstructed TOF may differ from zero but it should be the same for each event.
The result of the TOF distribution reconstructed 
for signals measured at z~=~150~mm is shown in the left panel of Fig.~\ref{figTOFResolution},
and the right panel shows the TOF resolution as a function of the position. 
The determined resolution is equal to about 125~ps($\sigma$) over the full length of 30~cm long detector.
It is important to stress that the TOF resolution includes contribution from the spread of the time of interaction 
due to the unknown depth of interaction of the gamma quantum 
within 19~mm thick scintillators (see Fig.~\ref{figSetup})
and due to the size of the source (3~mm in diameter). These two effects together 
cause a spread of about 27~ps($\sigma$).

\section{Summary}

A method enabling reconstruction of hit time and hit position of gamma 
quanta in scintillator detectors was described and validated based on the experimental
data collected with the double-module prototype of the J-PET detector. 
The method is based on a comparison of measured signal probed in the voltage 
or time domains with synchronized model signals from the library. 
The hit time and hit position are reconstructed as these which correspond to the signal 
from the library which is most similar to the measured signal. 
A measure of similarity is defined as the distance between points representing 
the measured- and model-signal in the multi-dimensional measurement space. 
In order to compare a measured signal (from the diagnosis of the patient) 
with the model signal from the synchronized library, minimization of the distance between points 
representing these signals is performed as a function of relative time between them. 
The relative time resultant from the minimization is taken as the time at which gamma 
quantum interacted in the detector.  Such choice of the value of time of interaction constitutes 
one of the crucial ideas of the described reconstruction method, 
and ensures that the difference between times of interactions reconstructed in different 
detectors for the quanta from the same annihilation process, 
correspond to the true difference (TOF) between times of arrival of these quanta to the detectors.
The novelty of the method lies also in synchronization of the model signals 
in a way enabling determination of TOF and in a manner of determining 
the time of the interaction of gamma quantum in detectors. In the article an exemplary procedure for generating 
a library of model signals was also presented which is based on scanning the scintillator 
strip with a collimated beam of annihilation quanta with profile smaller than the spatial resolution 
required for the hit-position reconstruction. 

The introduced method was validated by means of the experimental data collected by 
the double strip prototype of the J-PET detector built   
out from 
plastic scintillator strips
with dimensions of 5~mm x 19~mm x 300~mm 
read out at both sides by photomultipliers.
The strips were irradiated by the annihilation quanta from the 
$^{22}Na$ source placed in the middle of a lead collimator.
A library of model signals was created 
using a dedicated electro-mechanical system permitting to 
move the collimator along the scintillators 
in a way synchronized with the data acquisition system. 
Signals from photomultipliers were sampled with 100 ps intervals by means of the Serial Data Analyzer.
Applying the method introduced in this article a spatial resolution of about 1.3~cm ($\sigma$) for the hit-position reconstruction
and TOF resolution of about 125~ps ($\sigma$) were established.   

The obtained result for the TOF resolution for the detector of 30~cm length 
is better by about a factor of two with respect to the current TOF-PET tomographs characterized by typical 
field of views of about 16~cm and TOF resolution of about 230~ps ($\sigma$)~\cite{GEminiUJ}.

The result presented in this article can still be improved in the future
by more elaborated method of the determination of the onset of the signals 
used for the synchronization of the library of model signals 
(e.g. by utilising more than one value of time determined at different threshold levels)   
and by application of the measure of similarities which would account for the possible correlations between the times measured at different thresholds.

\section{Acknowledgements}

We acknowledge technical and administrative support by T. Gucwa-Ry\'{s}, A. Heczko,
M. Kajetanowicz, G. Konopka-Cupia\l{}, W. Migda\l{}, K. Wo\l{}ek and the financial support 
by the Polish National Center for Development and Research through grant No. INNOTECH-K1/IN1/64/159174/NCBR/12, 
the Foundation for Polish Science through MPD programme, 
the EU, MSHE Grant No. POIG.02.03.00-161 00-013/09, and Doctus - the Lesser Poland PhD Scholarship Fund.

%

%
%
%
\end{document}